\documentclass[a4paper,12pt]{article}
\usepackage{amsmath}
\usepackage{amssymb}
\usepackage{listings}
\usepackage{comment}

\usepackage[pdftex]{graphicx}
\newcommand {\Vect} [1] {{\bf #1}}
\newcommand {\F} {\Vect{F}}
\newcommand {\K} {\Vect{K}}
\newcommand {\rr} {\Vect{r}}
\newcommand {\vv} {\Vect{v}}
\newcommand {\be} [1]   {\begin{equation}\label{#1}}
\newcommand {\ee}       {\end{equation}}

\newcommand {\dt} {{\rm dt}}

\newcommand {\eq} [1]   {(\ref{#1})}

\begin{document}
\bigskip
\begin{center}
{\LARGE
On angular momentum balance for particle systems with periodic boundary conditions}
 \\ [4mm]
\large
Vitaly A. Kuzkin\footnote{kuzkinva@gmail.com}
 \\ [4mm]
 Institute for Problems in Mechanical Engineering RAS \\ [4mm]
  Saint Petersburg State Polytechnical University \\ [4mm]
\end{center}
\bigskip
\bigskip
\begin{abstract}
The well-known issue with the absence of conservation of angular momentum in classical particle systems with periodic boundary conditions is addressed. It is shown that conventional theory based on Noether's theorem fails to explain the simplest possible example, notably jumps of angular momentum in the case of single particle moving in a periodic cell. It is suggested to consider the periodic cell as an {\it open system}, exchanging mass, momentum, angular momentum, and energy with surrounding cells. Then the behavior of the cell is described by balance laws rather than conservation laws. It is shown using the law of angular momentum balance that the variation of the angular momentum in systems with periodic boundary conditions is a consequence of (i) the non-zero flux of angular momentum through the boundaries and (ii) torque acting on the cell due to the interactions between particles in the cell with images in the neighboring cells. Two simple examples demonstrating both phenomena are presented.
\end{abstract}

\section{Introduction}
Periodic boundary conditions are widely used in particle-based simulation techniques such as molecular dynamics~\cite{Hoover_MD}, particle dynamics~\cite{Krivtsov}, discrete element method~\cite{Cundal_Strack}, smoothed particles hydrodynamics~\cite{Lucy, Monaghan, Hoover_SPH}, dissipative particle dynamics~\cite{DPD}, etc. The total mass, momentum, and total energy\footnote{The energy is conserved if the particles interact via potential forces.} of the particle system with periodic boundary conditions are exactly conserved. Therefore the system is sometimes considered as isolated~\cite{White}. At the same time it is well known that the angular momentum of the particle system with periodic boundary conditions in general is not conserved. In particular, in the book~\cite{Hoover_MD} it is noted that ``periodic boundaries have the unusual property that they are inconsistent with conservation of angular momentum''. The conventional explanation of this fact is based on the Noether's theorem~\cite{Noether}. The theorem states that the conservation of angular momentum of an {\it isolated system} follows from the rotational invariance of its Largangian. Therefore it is often emphasized~(without rigorous proof) that the absence of conservation of angular momentum in particle systems under periodic boundary conditions is caused by the absence of rotational invariance of the Lagrangian~\cite{Rapaport, Haile, Frenkel_book, Frenkel}. However this approach fails to explain the simplest possible example, considered in the book by Hoover~\cite{Hoover_MD}. In the book~\cite{Hoover_MD} it is shown that the angular momentum for periodic cell containing single particle is not conserved.  Note that in this case the  Lagrangian of the particle is constant and therefore rotationally invariant. Then the absence of conservation of angular momentum is in contradiction with Noether's theorem. This ambiguity is discussed in more details in the next paragraph. Therefore the Noether's theorem is insufficient for explanation of the absence of conservation of angular momentum in particle systems with periodic boundary conditions, because there is at least one counterexample. Furthermore the theorem does not describe the evolution of the angular momentum.

In the present paper the periodic cell is considered as an {\it open system} exchanging mass, momentum, angular momentum, and energy through the boundary. Then the behavior of the system should be described by balance laws~\cite{Noll, Eringen, Zhilin_APM, Zhilin} rather than conservation laws. The law of angular momentum balance is formulated for the periodic cell. It is demonstrated using two simple examples that the change of angular momentum of the periodic cell is a consequence of (i) flux of angular momentum through the boundary and (ii) torque acting on the cell due to interaction with image particles. Therefore the law of angular momentum balance provides simple rational explanation for change of angular momentum in particle systems with periodic boundary conditions.

\section{The drawback of the conventional explanation based on Noether's theorem}
\begin{figure}[!htb]
\centering \includegraphics[scale=0.45]{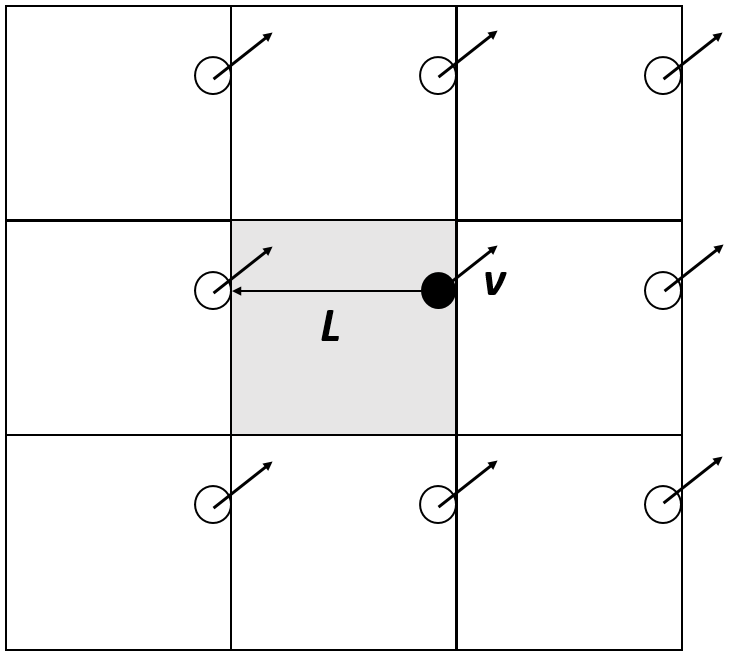}
\caption{Particle (black circle) in the cell with periodic boundary conditions. Unfilled circles correspond to images of the particle.}
\label{one part}
\end{figure}

Let us repeat the simplest example, given in the book~\cite{Hoover_MD}. Consider the motion of particle in the periodic cell~(see figure~\ref{one part}). Assume that the particle does not interact with its images in the neighboring cells. In this case the particle and all its images move with constant velocity~$\Vect{v}$.
Assume that the particle crosses periodic boundary at the moment~$t=t_{*}$.
At the same moment of time particle's image crosses the opposite boundary of the periodic cell.
Let us define the {\it angular momentum of the cell}~\Vect{K} as the total angular momentum of all particles
occupying the cell at the given moment of time. Then~$\Vect{K}$   is equal to the angular momentum of
the particle~$\Vect{r}\times m \Vect{v}$ until it does not leave the cell. When the particle leaves the
cell and its image enters the cell, the angular momentum of the cell~$\Vect{K}$  becomes equal
to~$\Vect{r}_{im}\times m_{im} \Vect{v}_{im}$. Therefore the difference between the angular momentum
of the cell~$\Vect{K}$ at~$t=t_{*}-\Delta t$   and $t=t_{*}+\Delta t$  is
\be{1}
\Vect{K}(t_{*}+\Delta t) - \Vect{K}(t_{*}-\Delta t) = m_{im} \Vect{r}_{im}\times \Vect{v}_{im} - m \Vect{r}\times \Vect{v},
\ee
where~$m, m_{im}, \Vect{r}, \Vect{r}_{im}, \Vect{v}, \Vect{v}_{im}$  are masses, radius-vectors, and velocities of the particle and its image. By the definition of periodic boundary conditions, the following relations are satisfied
\be{2}
   m=m_{im},~~~\Vect{r}=\Vect{r}_{im}-\Vect{L},~~~\Vect{v}=\Vect{v}_{im},
\ee
where~\Vect{L} is the constant vector characterizing the difference between coordinates of the particle and its image due
to periodic boundary conditions~(see figure~\ref{one part}). Substituting formula~\eq{2} into formula~\eq{1} one obtains
\be{3}
\Vect{K}(t_{*}+\Delta t) - \Vect{K}(t_{*}-\Delta t) = m \Vect{L}\times\Vect{v},
\ee
From~\eq{3} it follows that the angular momentum of the cell is conserved only if vectors~$\Vect{L}$ and~$\Vect{v}$   are parallel, i.e. the velocity of the particle is perpendicular to the side of the periodic cell. Thus, in general, the angular momentum of the cell is not conserved. This fact is usually refereed to  as the absence of conservation of angular momentum in particle systems with periodic boundary conditions. It is stated that the absence of conservation of angular momentum follows from the absence of rotational invariance of system's Largangian~\cite{Rapaport, Haile, Frenkel_book, Frenkel}. The rotational invariance is required by the Noether's theorem~\cite{Noether}.  However the straightforward application of the theorem to the considered example leads to the contradiction. The Lagrangian of the particle,~$m\Vect{v}^2/2$, is invariant to rotational transformations, while the angular momentum of the cell~$\Vect{K}$ is not conserved. This ambiguity is caused by the fact that the Noether's theorem is formulated for {\it isolated} systems~\cite{Noether}.  As we distinguish between particle and its images, then the periodic cell contains different particles at different moments of time. Therefore the cell is an {\it open system} exchanging mass, momentum, angular momentum, and energy with neighboring cells. Thus Noether's theorem can not be used for explanation of the absence of conservation of angular momentum in periodic systems.

\section{Balance of angular momentum for periodic cell considered as an open system}
Consider the periodic cell as an {\it open system}.  The behavior of open systems is described by the balance laws~\cite{Noll, Eringen, Zhilin_APM, Zhilin} rather than conservation
laws. The law of angular momentum balance for the periodic cell has the form
\be{LAMB}
\dot{\Vect{K}} = \Vect{T} + \Vect{Q}.
\ee
The first term~$\Vect{T}$ in the right side of equation~\eq{LAMB} is the total torque acting on the cell due to the interaction of particles in the cell with image particles in the neighboring cells. The second term~$\Vect{Q}$ is the flux of angular momentum into and out from the cell caused by the particles leaving and entering the cell through the periodic boundaries. The example considered in the previous paragraph shows that the angular momentum of the cell~$\K(t)$ could be discontinuous function of time~(see equation~\eq{3}). In this case it could be convenient to avoid the derivative in equation~\eq{LAMB}. Let us rewrite equation~\eq{LAMB} for system of material points
interacting via pair forces~$\F_{ij}$. Assume that at time~$t$ the cell is occupied by the set of particles~$\Lambda(t)$ and the neighboring cells are occupied by the set of particles~$\Lambda_{im}(t)$. Consider moments of time~$t \in [t_{*}; t_{*}+\Delta t]$. Assume that the set of particles~$\Lambda_{+}$ enter the cell and the set of particles~$\Lambda_{-}$ leave the cell during this interval.
Then the equation~\eq{LAMB} can be rewritten as
\be{LAMB_int}
\begin{array}{l}
\displaystyle \K(t_{*}+ \Delta t) - \K(t_{*}) = \int_{t_{*}}^{t_{*}+\Delta t} \left(\Vect{T}  + \Vect{Q} \right) \dt, \qquad \K(t) = \sum_{i \in \Lambda(t)} \rr_i \times m \vv_i \\[4mm]
\displaystyle \Vect{T}(t) = \sum_{i \in \Lambda(t), j \in \Lambda_{im}(t)} \rr_i \times \F_{ij},\qquad  \int_{t_{*}}^{t_{*}+\Delta t} \Vect{Q} \dt =  \sum_{i \in \Lambda_{+}} \rr_i \times m\vv_{i} - \sum_{i \in \Lambda_{-}} \rr_i \times m\vv_{i},
\end{array}
\ee
The equation~\eq{LAMB_int} shows that the change of angular momentum of the periodic cell is caused by (i) flux of angular momentum through the boundaries and (ii) torque acting on the cell. Two simple examples demonstrating this idea are given below.

\subsection{Example 1: Flux of angular momentum through the periodic boundaries}

Let us revisit the example discussed above. Consider single particle in the periodic cell. In this case the torque~$\Vect{T}$ acting on the cell is equal to zero as there are no interactions between the particle and its neighbors.
However the flux of angular momentum through the boundaries in general is not equal to zero. When the particle crosses the periodic boundary angular
momentum~$\Vect{r}\times m \Vect{v}$  flow out from the cell. At the same
time the image particle enters the cell from the opposite side and therefore its angular momentum~$ \Vect{r}_{im}\times m_{im}\Vect{v}_{im}$
flow into the cell. Then according to formula~\eq{LAMB_int} the flux of angular momentum
during the interval~$[t_{*} - \Delta t; t_{*} + \Delta t]$ has the form:
\be{}
  \int_{t_{*} - \Delta t}^{t_{*}+\Delta t} \Vect{Q} \dt =  \rr \times m\vv - \rr_{im} \times m\vv_{im} = m \Vect{L}\times\Vect{v}.
\ee
One can see that the right side of the given formula exactly coincides with the right side of formula~\eq{1}, describing the change of angular momentum during the interval~$[t_{*} - \Delta t; t_{*} + \Delta t]$. Therefore in the given example the absence of conservation of the angular momentum is a direct consequence of non-zero flux of angular momentum through the boundary.

\subsection{Example 2: Torque acting on the cell}
\begin{figure}[!htb]
\centering \includegraphics[scale=0.45]{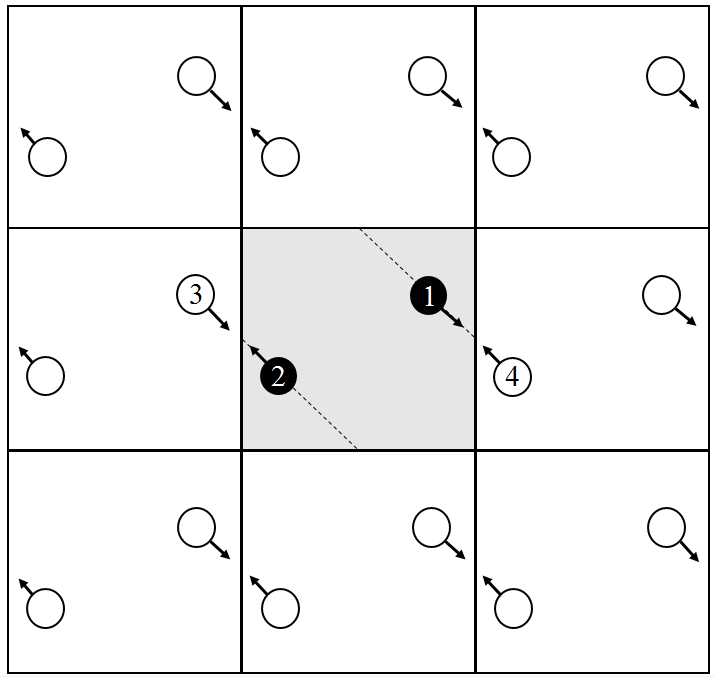}
\caption{Two particles in the cell with periodic boundary conditions. Unfilled circles correspond to images of the particles. The dashed lines show the trajectories of the particles.}
\label{two_particles}
\end{figure}

According to equation~\eq{LAMB} the angular momentum of the cell may change even in the absence of flux through the boundaries. This fact is emphasized, for example, in paper~\cite{Monaghan_swim}. The simplest example demonstrating this phenomenon is given below.
Consider two particles moving in the periodic cell along the dashed lines,
shown in figure~\ref{two_particles}. Assume that the interactions are purely repulsive and short-ranged, so that the particles
in one cell does not interact with each other. Then the only action on the cell is due to interaction with
image particles in neighboring cells. Then equation~\eq{LAMB} takes the form
\be{LAMB_2part}
  \dot{\K} = \rr_1 \times \F_{14} + \rr_2 \times \F_{23} = (\rr_1 - \rr_2)  \times \F_{14}.
\ee
Here we used the relation~$\F_{14} = \F_{32} = -\F_{23}$ following from Newton's third law and periodicity of the system.
If the distances~$r_{14}$ and~$r_{13}$ are larger than the cut-off radius, then~$\F_{14} = 0$ and the angular momentum is conserved. If the particles are closer than the cut-off distance then the angular momentum changes. The rate of change of angular momentum is given by formula~\eq{LAMB_2part}.
After the collision velocities of the particles and angular momentum of the cell~$\K$ change sign.
In the case of hard sphere interactions~$\K(t)$ is a peace-wise constant periodic function that takes two values: $\K_0$ before the collision and~$-\K_0$ after the collision.

Thus in the present example the change of angular momentum is a consequence of non-zero torque acting on the cell.
\section{Conclusions}
The absence of conservation of angular momentum in particle systems with periodic boundary conditions was considered. We have shown that the conventional explanation of this phenomenon based on rotational invariance  fails to explain the simplest possible example~(single particle moving in the periodic cell). It was suggested to consider the periodic cell as an {\it open system} and describe the behavior of the cell by balance laws rather than conservation laws. In particular the law of angular momentum balance for the periodic cell containing interacting particles was formulated. The law clearly shows two physical mechanisms that causes the change of angular momentum in periodic systems: (i) flux of angular momentum through the boundaries and (ii) torque acting on the cell. The flux is typical for liquids and gases and leads to discontinuous change of angular momentum as particles leave and enter the cell. For solids the second mechanism usually dominates. Two simple examples demonstrating both mechanisms were presented and analyzed. Thus the consideration of the periodic cell as an {\it open system} provides simple, rational explanation for the absence of conservation of angular momentum for particle systems with periodic boundary conditions.

The author is deeply grateful to D. Frenkel, W.G. Hoover, E.A. Ivanova, R.D. James, and A.M. Krivtsov for useful discussions.

\end{document}